\documentclass[preprint]{aastex}

\begin{document}

\title{\bf {RW Dor: A shallow contact binary with new orbital period investigation}}

\author{T. Sarotsakulchai\altaffilmark{1,2,3},
S.-B. Qian\altaffilmark{1,2,4,5},
B. Soonthornthum\altaffilmark{3},
E. Fern$\acute{a}$ndez Laj$\acute{u}$s\altaffilmark{6,7},\\
N.-P. Liu\altaffilmark{1,4,5},
X. Zhou\altaffilmark{1,4,5},
J. Zhang\altaffilmark{1,4,5},
W.-P. Liao\altaffilmark{1,4,5},\\
D. E. Reichart\altaffilmark{8},
J. B. Haislip\altaffilmark{8},
V. V. Kouprianov\altaffilmark{8} and
S. Poshyachinda\altaffilmark{3}}

\altaffiltext{1}{Yunnan Observatories, Chinese Academy of Sciences, 650216 Kunming, China} \email{huangbinghe@ynao.ac.cn}
\altaffiltext{2}{University of Chinese Academy of Sciences, 19 A Yuquan Rd., Shijingshan, 100049 Beijing, China}
\altaffiltext{3}{National Astronomical Research Institute of Thailand, Ministry of Science and Technology, Bangkok, Thailand}
\altaffiltext{4}{Key Laboratory of the Structure and Evolution of Celestial Objects, Chinese Academy of Sciences, 650216 Kunming, China}
\altaffiltext{5}{Center for Astronomical Mega-Science, Chinese Academy of Sciences, 20A Datun Rd., Chaoyang District, Beijing, 100012, China}
\altaffiltext{6}{Facultad de Ciencias Astron$\acute{o}$micas y Geof$\acute{i}$sicas - UNLP, 1900 La Plata, Buenos Aires, Argentina}
\altaffiltext{7}{Instituto de Astrofisica de La Plata - CONICET/UNLP, Argentina}
\altaffiltext{8}{Department of Physics and Astronomy, University of North Carolina, CB \#3255, Chapel Hill, NC 27599, USA}

\begin{abstract}
New CCD photometric light curves of short period (P=0.285d) eclipsing binary RW Dor are presented. The observations performed with the PROMPT-8 robotic telescope at CTIO in Chile from March 2015 to March 2017. The other eclipse timings were obtained from the 2.15-m JS telescope at CASLEO, San Juan, Argentina in December 2011. By light-curve analysis, it is found that RW Dor is a W-type shallow contact binary with a fill-out factor $f \sim 11\%$ and high mass ratio $q \sim 1.587$ (1/q = 0.63), where the hotter component is the less massive one ($M_1 \sim 0.52M_{\odot}$ and $M_2 \sim 0.82M_{\odot}$). For orbital period investigation, the new fifteen eclipse times and those in previous published were compiled. Based on $O-C$ analysis with very weak evidence suggests that a long-term period decrease with a rate of $\mathrm{d}P/\mathrm{d}t = -9.61\times10^{-9}$ d $\textrm{yr}^{-1}$ is superimposed on a cyclic variation ($A_3$ = 0.0054 days and $P_3$ = 49.9 yrs). The long-term period decrease can be interpreted as mass transfer from the more massive component to the less massive one or combine with the angular momentum loss (AML) via magnetic braking. In addition, with the marginal contact phase, high mass ratio (1/q $>$ 0.4) and the long-term period decrease, all suggest that RW Dor is a newly formed contact binary via a Case A mass transfer and it will evolve into a deeper normal contact binary. If the cyclic change is correct, the light-travel time effect via the presence of a cool third body will be more plausible to explain for this.
\end{abstract}


\section{Introduction}

RW Dor (HD 269320, HIP 24763) is an important binary for studying the formation and evolution at the beginning or newly formed contact binary after the common convective envelope has been formed and in the transition either from detached or semi-detached into the contact phase {\bf because there is a few or very rare contact binaries that have been found to be newly formed or at the beginning of the contact phase that they will evolve into a normal contact binary stars when their mass ratio become higher e.g. RV Psc (He \& Qian 2009) and V524 Mon (He et al. 2012)}. RW Dor is a short period W UMa-type binary with an orbital period of 0.285\,days {\bf that is very close to a new period distribution of EW-type contact binaries with a peak of 0.29\,days published by Qian et al. (2017), indicating that RW Dor is a typical W UMa-type contact binary that is in agreement with the report (Qian et al. 2017)}. The system is near the Large Magellanic Cloud (LMC), but not a member of LMC as pointed out by Russo et al. (1984) and Marino et al. (2007). RW Dor was discovered as a variable star by Leavitt (1908) and later classified as a W UMa-type eclipsing binary by Hertzsprung (1925). The first spectral classification of the variable was made by Cannon (1921) who gave a spectral type of K5. This was subsequently confirmed by McLaughlin (1927) who also classified the system as a late-type eclipsing binary with the same spectral type.

The first photographic times of light minimum were reported by Hertzsprung (1928), who gave an orbital period of 0.143 days. Later, a lot of eclipse times were obtained by many authors (e.g. Russo et al. 1984; Marton et al. 1989; Kaluzny \& Caillault 1989; Ogloza \& Zakrzewski 2004; Marino et al. 2007) and the linear ephemeris of the binary was also corrected. The complete light curves were analysed with Wilson-Devinney method (Wilson \& Devinney 1971) by Marton et al. (1989) and Kaluzny \& Caillault (1989) who determined photometric elements independently. Those solutions indicated that RW Dor belongs to a W-subtype contact binary with components are not in poor thermal contact which predicted by the thermal relaxation oscillations theory (Lucy \& Wilson 1979). Additionally, it was found that the light curves of RW Dor exhibited a significant difference in depths of the minima and showed variation. Marton et al. (1989) explained the asymmetry in the light curves as a hot spot on the cooler and more massive component located near the neck connecting the stars, while Kaluzny \& Caillault (1989) reported that their light curves were only marginally asymmetric and did not show any scatter more than the observational errors.

The first radial-velocity measurements of RW Dor were published by Hilditch et al. (1992) with the 3.9-m telescope of Anglo-Australian Observatory. They found that RW Dor was composed of two K1 type stars and determined a spectroscopic mass ratio of $q_{sp}=0.68$. By combining the photometric solutions given by Kaluzny \& Caillault (1989), Hilditch et al. (1992) derived absolute parameters of the binary system. They confirmed that RW Dor is a W-subtype contact binary. The other radial-velocity curves were obtained by Duerbeck \& Rucinski (2007) who determined a spectroscopic mass ratio of $q_{sp}=0.63$. This is close to that derived by Hilditch et al. (1992). However, they found $V_{0}=25$ km/s that is quite smaller than the one ($V_{0}=66.5$ km/s) given by Hilditch et al. (1992). Marino et al. (2007) reported that some light curves of RW Dor exhibited asymmetry, which were similar to the results published by Marton et al. (1989). They also re-computed the spectroscopic mass ratio by using the radial velocities given by Hilditch et al. (1992) and Duerbeck \& Rucinski (2007). All eclipse times were also re-analyzed and they found a secular decrease in orbital period with a rate of $\Delta{P}/P \sim -6.3\times10^{-11}$. However, no light curve variation and third light in the system were found in recent publication (e.g. Deb $\&$ Singh 2011).

In this paper, we examine the variations of the light curve and determine new photometric solutions based on our CCD observations. Then, we compare our results to those from the other investigators. The orbital period changes are re-investigated with new eclipse times together with the others collected from the literature. Finally, the formation and the evolutionary state of the system, as well as the probability of the additional companion orbiting around the contact binary are all discussed.

\begin{table}
\scriptsize
\caption{Coordinates of RW Dor, the comparison and the check stars.}
\begin{center}
\begin{tabular}{llcccccccc}\hline\hline
Targets & Names & $\alpha_{2000}$ & $\delta_{2000}$&$V$&$R$&$J$&$H$&$K$\\
\hline
Variable star & RW Dor &$05^{\textrm{h}}18^{\textrm{m}}32^{\textrm{s}}.5$ & $-68^\circ13^{\prime}32^{\prime\prime}.7$ &11.16&8.66&9.282&8.781&8.709\\
The comparison& GSC0916200441 &$05^{\textrm{h}}18^{\textrm{m}}43^{\textrm{s}}.6$ & $-68^\circ07^{\prime}33^{\prime\prime}.7$ &12.25&11.89&11.916&11.859&11.863\\
The check     & J05175264-6813241 &$05^{\textrm{h}}17^{\textrm{m}}52^{\textrm{s}}.6$ & $-68^\circ13^{\prime}24^{\prime\prime}.1$ &12.33&11.27&10.782&10.285&10.169\\
\hline
\end{tabular}
\end{center}
\end{table}

\section{CCD photometric observations}

\begin{figure}
\begin{center}
\includegraphics[angle=0,scale=0.5]{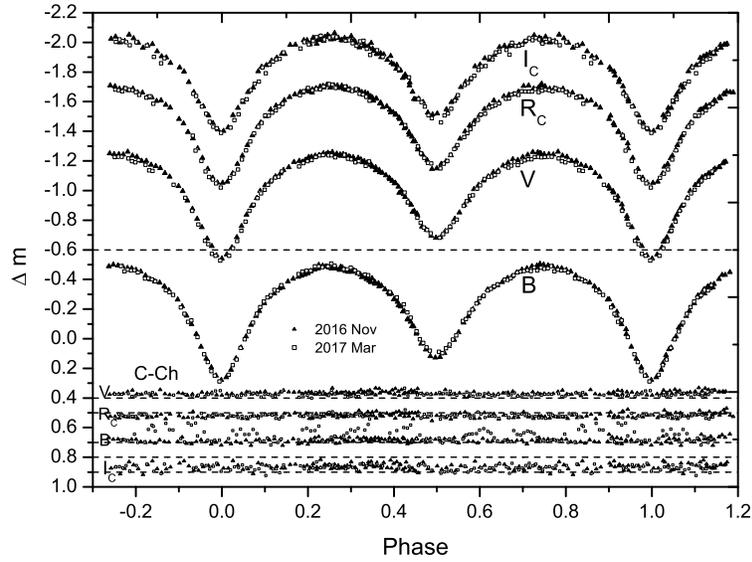}
\caption{The two sets for second observation with complete light curves in $B, V, R_{c}, I_{c}$ bands were obtained with the 0.6-m telescope at CTIO during November 2016 (dark triangles) and March 2017 (open squares). The differential magnitude between the comparison and check stars in $BVR_{c}I_{c}$ bands (C-Ch) are plotted at the bottom of the figure and they were used to calibrate all data sets in each band.}
\end{center}
\end{figure}

The first observations in V-band with two new times of light minimum were obtained by using the 2.15-m "Jorge Sahade" (JS) telescope at Complejo Astronomico El Leoncito (CASLEO), San Juan, Argentina during 9, 11 and 13 December 2011. During the observations, a VersArray 1300B camera was attached to the 2.15-m telescope with exposure 1095, 1300 and 1000 images for 2011 December 9, 11 and 13, respectively. The second observations in $BV(RI)_{C}$ bands were obtained from March 2015 to March 2017 by using the back illuminated Apogee F42 2048$\times$2048 CCD photometric system, attached to the 0.60-m Cassegrain reflecting telescope of PROMPT-8\footnote{PROMPT-8 is the Thai Southern Hemisphere Telescope (TST), operated in collaboration between National Astronomical Research Institute of Thailand (NARIT) and the University of North Carolina (UNC) at Chapel Hill in a part of the UNC-led PROMPT project, http://skynet.unc.edu.} robotic telescope, which is located at the Cerro Tololo Inter-American Observatory (CTIO) in Chile. It also provided nightly calibration images, including bias, dark, and flat-field images (Layden \& Broderick 2010). The CCD reduction and aperture photometry were done with standard procedure packages of IRAF\footnote{The Image Reduction and Analysis Facility (IRAF), http://iraf.noao.edu.}. The comparison star (C) is GSC 09162-00441; $V$= 12.25, $J-H$= 0.057 (SIMBAD) and the check star (Ch) is 2MASS J05175264-6813241; $V$= 12.33, $J-H$ = 0.497 (SIMBAD). The coordinates of those targets are listed in Table 1 and the complete multi-color light curves for second observation in November 2016 and in March 2017 are displayed in dark triangles and open squares of Fig. 1, respectively. The data from Fig.1 are listed in Table 2 for 2016 and Table 3 for 2017. The comparison of the two sets of light curves are also plotted together in the figure, they overlap nearly indicating that the light curves have no any changes and they are quite symmetric. {\bf This means that the light curves for our observations at this time are stable, which are opposite to some results from previous investigators who found that the light curves showed asymmetry and varied with time}. The depths of the primary and the secondary minima of the two sets of light curves from Fig. 1 are displayed in Table 4 and no any significant changes were found there. {\bf This helps to confirm that the light curves have no variations or asymmetry from O'Connell effect (O'Connell 1951) caused by spot or magnetic activities which normally occur similar to the sun or late type stars e.g. spectral type F,G and K.}

\begin{table}
\scriptsize
\caption{CCD observations of RW Dor in November 2016}
\begin{center}
\begin{tabular}{lrlrlrlrlrlr}\hline\hline
HJD&$\Delta V$&HJD&$\Delta R$&HJD&$\Delta I$
&HJD&$\Delta V$&HJD&$\Delta R$&HJD&$\Delta I$\\
+2457700&&&&&&&&&&&\\
\hline
20.5558	&	-1.228	&	20.5562	&	-1.641	&	20.5566	&	-2.017	&	20.8129	&	-1.061	&	20.8134	&	-1.522	&	20.8138	&	-1.894	\\
20.5585	&	-1.233	&	20.5589	&	-1.664	&	20.5593	&	-1.991	&	20.8151	&	-1.090	&	20.8155	&	-1.531	&	20.8159	&	-1.906	\\
20.5612	&	-1.238	&	20.5616	&	-1.681	&	20.5619	&	-2.039	&	20.8172	&	-1.099	&	20.8177	&	-1.539	&	20.8180	&	-1.918	\\
20.5637	&	-1.250	&	20.5642	&	-1.676	&	20.5646	&	-2.037	&	20.8193	&	-1.116	&	20.8198	&	-1.557	&	20.8202	&	-1.928	\\
20.5667	&	-1.255	&	20.5672	&	-1.692	&	20.5675	&	-2.039	&	20.8213	&	-1.117	&	20.8218	&	-1.567	&	20.8223	&	-1.921	\\
20.5696	&	-1.260	&	20.5701	&	-1.686	&	20.5706	&	-2.023	&	20.8235	&	-1.147	&	20.8240	&	-1.580	&	20.8244	&	-1.949	\\
20.5727	&	-1.238	&	20.5731	&	-1.675	&	20.5735	&	-2.029	&	20.8267	&	-1.158	&	20.8272	&	-1.604	&	20.8276	&	-1.965	\\
20.5756	&	-1.234	&	20.5760	&	-1.688	&	20.5764	&	-2.012	&	20.8289	&	-1.183	&	20.8294	&	-1.633	&	20.8298	&	-1.987	\\
20.5784	&	-1.232	&	20.5789	&	-1.659	&	20.5793	&	-2.015	&	20.8307	&	-1.194	&	20.8312	&	-1.648	&	20.8316	&	-1.990	\\
20.5814	&	-1.225	&	20.5818	&	-1.656	&	20.5822	&	-2.042	&	21.5639	&	-1.271	&	21.5643	&	-1.709	&	21.5648	&	-2.039	\\
20.5842	&	-1.210	&	20.5845	&	-1.656	&	20.5849	&	-2.003	&	21.5668	&	-1.265	&	21.5673	&	-1.704	&	21.5677	&	-2.054	\\
20.5868	&	-1.220	&	20.5873	&	-1.656	&	20.5877	&	-1.966	&	21.5700	&	-1.265	&	21.5704	&	-1.687	&	21.5709	&	-2.015	\\
20.5897	&	-1.190	&	20.5902	&	-1.630	&	20.5907	&	-1.978	&	21.5729	&	-1.264	&	21.5735	&	-1.688	&	21.5739	&	-2.024	\\
20.5926	&	-1.175	&	20.5931	&	-1.613	&	20.5936	&	-1.973	&	21.5760	&	-1.282	&	21.5763	&	-1.677	&	21.5767	&	-2.067	\\
20.5957	&	-1.170	&	20.5961	&	-1.608	&	20.5964	&	-1.957	&	21.5789	&	-1.256	&	21.5793	&	-1.687	&	21.5798	&	-2.042	\\
20.5984	&	-1.137	&	20.5988	&	-1.601	&	20.5991	&	-1.921	&	21.5820	&	-1.249	&	21.5824	&	-1.663	&	21.5828	&	-2.012	\\
20.6012	&	-1.123	&	20.6017	&	-1.573	&	20.6021	&	-1.901	&	21.5851	&	-1.220	&	21.5856	&	-1.661	&	21.5860	&	-1.979	\\
20.6080	&	-1.077	&	20.6085	&	-1.507	&	20.6089	&	-1.861	&	21.5882	&	-1.220	&	21.5887	&	-1.629	&	21.5892	&	-1.994	\\
20.6110	&	-1.049	&	20.6115	&	-1.491	&	20.6119	&	-1.821	&	21.5914	&	-1.194	&	21.5919	&	-1.628	&	21.5923	&	-2.003	\\
20.6140	&	-0.984	&	20.6144	&	-1.463	&	20.6148	&	-1.815	&	21.5945	&	-1.184	&	21.5950	&	-1.605	&	21.5954	&	-1.954	\\
20.6170	&	-0.975	&	20.6175	&	-1.401	&	20.6179	&	-1.783	&	21.5975	&	-1.172	&	21.5980	&	-1.609	&	21.5984	&	-1.921	\\
20.6201	&	-0.927	&	20.6205	&	-1.373	&	20.6208	&	-1.709	&	21.6006	&	-1.138	&	21.6011	&	-1.579	&	21.6015	&	-1.913	\\
20.6229	&	-0.855	&	20.6233	&	-1.311	&	20.6237	&	-1.618	&	21.6036	&	-1.114	&	21.6040	&	-1.560	&	21.6043	&	-1.876	\\
20.6259	&	-0.845	&	20.6264	&	-1.256	&	20.6267	&	-1.613	&	21.6095	&	-1.050	&	21.6100	&	-1.483	&	21.6103	&	-1.846	\\
20.6291	&	-0.757	&	20.6295	&	-1.223	&	20.6298	&	-1.597	&	21.6124	&	-1.016	&	21.6129	&	-1.453	&	21.6133	&	-1.818	\\
20.6321	&	-0.735	&	20.6326	&	-1.184	&	20.6329	&	-1.526	&	21.6156	&	-0.952	&	21.6159	&	-1.394	&	21.6164	&	-1.746	\\
20.6352	&	-0.694	&	20.6357	&	-1.126	&	20.6361	&	-1.512	&	21.6188	&	-0.897	&	21.6192	&	-1.351	&	21.6197	&	-1.680	\\
20.6384	&	-0.682	&	20.6388	&	-1.130	&	20.6393	&	-1.520	&	21.6219	&	-0.828	&	21.6224	&	-1.285	&	21.6228	&	-1.615	\\
20.6414	&	-0.686	&	20.6419	&	-1.160	&	20.6423	&	-1.480	&	21.6250	&	-0.772	&	21.6255	&	-1.220	&	21.6259	&	-1.584	\\
20.6446	&	-0.727	&	20.6451	&	-1.177	&	20.6454	&	-1.548	&	21.6285	&	-0.696	&	21.6290	&	-1.145	&	21.6295	&	-1.522	\\
20.6477	&	-0.752	&	20.6483	&	-1.217	&	20.6486	&	-1.597	&	21.6317	&	-0.645	&	21.6322	&	-1.085	&	21.6325	&	-1.470	\\
20.6509	&	-0.809	&	20.6513	&	-1.268	&	20.6517	&	-1.624	&	21.6349	&	-0.583	&	21.6354	&	-1.055	&	21.6358	&	-1.427	\\
20.6540	&	-0.852	&	20.6545	&	-1.330	&	20.6549	&	-1.698	&	21.6384	&	-0.561	&	21.6389	&	-1.052	&	21.6394	&	-1.417	\\
20.6573	&	-0.902	&	20.6576	&	-1.379	&	20.6581	&	-1.733	&	21.6417	&	-0.590	&	21.6421	&	-1.059	&	21.6425	&	-1.441	\\
20.6603	&	-0.935	&	20.6608	&	-1.419	&	20.6612	&	-1.751	&	21.6450	&	-0.636	&	21.6455	&	-1.116	&	21.6458	&	-1.524	\\
20.6636	&	-1.010	&	20.6639	&	-1.450	&	20.6644	&	-1.838	&	21.6482	&	-0.706	&	21.6487	&	-1.179	&	21.6491	&	-1.573	\\
20.6668	&	-1.040	&	20.6673	&	-1.489	&	20.6677	&	-1.850	&	21.6515	&	-0.780	&	21.6519	&	-1.235	&	21.6524	&	-1.613	\\
20.6701	&	-1.068	&	20.6706	&	-1.530	&	20.6709	&	-1.892	&	21.6548	&	-0.857	&	21.6553	&	-1.306	&	21.6557	&	-1.680	\\
20.6732	&	-1.096	&	20.6737	&	-1.552	&	20.6741	&	-1.883	&	21.6582	&	-0.917	&	21.6586	&	-1.373	&	21.6591	&	-1.710	\\
20.6765	&	-1.123	&	20.6769	&	-1.589	&	20.6773	&	-1.915	&	21.6616	&	-0.970	&	21.6621	&	-1.430	&	21.6625	&	-1.819	\\
20.6797	&	-1.134	&	20.6802	&	-1.574	&	20.6806	&	-1.968	&	21.6637	&	-1.017	&	21.6642	&	-1.459	&	21.6646	&	-1.849	\\
20.6830	&	-1.168	&	20.6835	&	-1.616	&	20.6839	&	-1.978	&	21.6670	&	-1.058	&	21.6675	&	-1.501	&	21.6678	&	-1.862	\\
20.6863	&	-1.188	&	20.6868	&	-1.625	&	20.6870	&	-1.973	&	21.6703	&	-1.080	&	21.6708	&	-1.547	&	21.6713	&	-1.892	\\
20.6894	&	-1.212	&	20.6899	&	-1.632	&	20.6904	&	-1.978	&	21.6738	&	-1.116	&	21.6742	&	-1.572	&	21.6747	&	-1.911	\\
20.6928	&	-1.205	&	20.6932	&	-1.646	&	20.6936	&	-1.982	&	21.6773	&	-1.135	&	21.6778	&	-1.583	&	21.6782	&	-1.938	\\
20.6961	&	-1.227	&	20.6966	&	-1.685	&	20.6970	&	-2.002	&	21.6793	&	-1.159	&	21.6797	&	-1.610	&	21.6802	&	-1.936	\\
20.6994	&	-1.242	&	20.6998	&	-1.690	&	20.7002	&	-2.026	&	21.6827	&	-1.177	&	21.6832	&	-1.631	&	21.6837	&	-1.998	\\
20.7026	&	-1.260	&	20.7031	&	-1.669	&	20.7036	&	-2.048	&	21.6863	&	-1.213	&	21.6868	&	-1.649	&	21.6872	&	-2.000	\\
20.7060	&	-1.259	&	20.7065	&	-1.676	&	20.7069	&	-2.031	&	21.6898	&	-1.221	&	21.6903	&	-1.664	&	21.6906	&	-1.957	\\
20.7080	&	-1.251	&	20.7085	&	-1.700	&	20.7089	&	-2.006	&	21.6917	&	-1.214	&	21.6922	&	-1.663	&	21.6926	&	-2.030	\\
20.7538	&	-1.013	&	20.7542	&	-1.485	&	20.7546	&	-1.814	&	21.6938	&	-1.228	&	21.6944	&	-1.675	&	21.6948	&	-2.030	\\
20.7563	&	-0.987	&	20.7568	&	-1.419	&	20.7573	&	-1.777	&	21.6959	&	-1.247	&	21.6964	&	-1.668	&	21.6969	&	-2.002	\\
20.7591	&	-0.960	&	20.7596	&	-1.378	&	20.7601	&	-1.749	&	21.6979	&	-1.242	&	21.6985	&	-1.683	&	21.6989	&	-2.047	\\
20.7617	&	-0.906	&	20.7621	&	-1.355	&	20.7625	&	-1.703	&	21.7003	&	-1.254	&	21.7008	&	-1.691	&	21.7013	&	-2.032	\\
20.7651	&	-0.828	&	20.7656	&	-1.299	&	20.7660	&	-1.615	&	21.7028	&	-1.260	&	21.7032	&	-1.685	&	21.7037	&	-2.031	\\
20.7895	&	-0.656	&	20.7900	&	-1.141	&	20.7904	&	-1.495	&	21.7050	&	-1.261	&	21.7055	&	-1.705	&	21.7059	&	-2.024	\\
20.7915	&	-0.676	&	20.7920	&	-1.153	&	20.7923	&	-1.530	&	21.7075	&	-1.263	&	21.7081	&	-1.713	&	21.7084	&	-2.065	\\
20.7937	&	-0.724	&	20.7942	&	-1.199	&	20.7946	&	-1.558	&	21.7099	&	-1.265	&	21.7104	&	-1.694	&	21.7108	&	-2.035	\\
20.7970	&	-0.808	&	20.7975	&	-1.254	&	20.7980	&	-1.643	&	21.7122	&	-1.265	&	21.7127	&	-1.709	&	21.7131	&	-2.078	\\
20.8002	&	-0.869	&	20.8007	&	-1.330	&	20.8010	&	-1.710	&	21.7145	&	-1.268	&	21.7150	&	-1.709	&	21.7155	&	-2.047	\\
\hline
\end{tabular}
\end{center}
\end{table}

\begin{table}
\scriptsize
\caption{CCD observations of RW Dor in March 2017}
\begin{center}
\begin{tabular}{lrlrlrlrlrlr}\hline\hline
HJD&$\Delta V$&HJD&$\Delta R$&HJD&$\Delta I$
&HJD&$\Delta V$&HJD&$\Delta R$&HJD&$\Delta I$\\
+2457800&&&&&&&&&&&\\
\hline
26.5331	&	-0.871	&	26.5335	&	-1.338	&	26.5339	&	-1.683	&	26.6569	&	-1.095	&	26.6433	&	-1.626	&	26.6699	&	-1.719	\\
26.5353	&	-0.837	&	26.5356	&	-1.290	&	26.5359	&	-1.663	&	26.6645	&	-1.007	&	26.6695	&	-1.410	&	26.6717	&	-1.703	\\
26.5374	&	-0.806	&	26.5378	&	-1.261	&	26.5381	&	-1.633	&	26.6691	&	-0.919	&	26.6713	&	-1.310	&	27.5323	&	-1.642	\\
26.5396	&	-0.789	&	26.5401	&	-1.227	&	26.5404	&	-1.575	&	26.6731	&	-0.826	&	27.5319	&	-1.251	&	27.5350	&	-1.593	\\
26.5420	&	-0.766	&	26.5424	&	-1.223	&	26.5428	&	-1.553	&	27.5342	&	-0.725	&	27.5346	&	-1.179	&	27.5373	&	-1.540	\\
26.5479	&	-0.737	&	26.5484	&	-1.190	&	26.5487	&	-1.525	&	27.5365	&	-0.703	&	27.5370	&	-1.142	&	27.5397	&	-1.506	\\
26.5503	&	-0.761	&	26.5507	&	-1.221	&	26.5511	&	-1.572	&	27.5388	&	-0.638	&	27.5393	&	-1.124	&	27.5420	&	-1.497	\\
26.5526	&	-0.781	&	26.5530	&	-1.240	&	26.5534	&	-1.586	&	27.5412	&	-0.625	&	27.5416	&	-1.098	&	27.5450	&	-1.464	\\
26.5548	&	-0.811	&	26.5552	&	-1.282	&	26.5556	&	-1.617	&	27.5443	&	-0.593	&	27.5447	&	-1.070	&	27.5511	&	-1.499	\\
26.5571	&	-0.843	&	26.5575	&	-1.320	&	26.5579	&	-1.671	&	27.5503	&	-0.624	&	27.5507	&	-1.136	&	27.5537	&	-1.551	\\
26.5593	&	-0.859	&	26.5596	&	-1.347	&	26.5600	&	-1.709	&	27.5529	&	-0.675	&	27.5533	&	-1.164	&	27.5564	&	-1.589	\\
26.5639	&	-0.951	&	26.5642	&	-1.402	&	26.5646	&	-1.778	&	27.5556	&	-0.724	&	27.5561	&	-1.192	&	27.5591	&	-1.681	\\
26.5659	&	-0.983	&	26.5662	&	-1.428	&	26.5666	&	-1.791	&	27.5583	&	-0.794	&	27.5588	&	-1.259	&	27.5619	&	-1.691	\\
26.5680	&	-1.020	&	26.5684	&	-1.456	&	26.5688	&	-1.823	&	27.5611	&	-0.855	&	27.5615	&	-1.322	&	27.5646	&	-1.753	\\
26.5700	&	-1.023	&	26.5705	&	-1.490	&	26.5708	&	-1.840	&	27.5637	&	-0.929	&	27.5642	&	-1.381	&	27.5672	&	-1.795	\\
26.5728	&	-1.068	&	26.5732	&	-1.518	&	26.5736	&	-1.879	&	27.5665	&	-0.969	&	27.5668	&	-1.436	&	27.5699	&	-1.811	\\
26.5748	&	-1.091	&	26.5753	&	-1.553	&	26.5756	&	-1.903	&	27.5691	&	-1.014	&	27.5695	&	-1.481	&	27.5722	&	-1.907	\\
26.5776	&	-1.134	&	26.5781	&	-1.558	&	26.5784	&	-1.906	&	27.5714	&	-1.046	&	27.5719	&	-1.507	&	27.5746	&	-1.867	\\
26.5804	&	-1.148	&	26.5808	&	-1.600	&	26.5812	&	-1.965	&	27.5738	&	-1.071	&	27.5742	&	-1.557	&	27.5769	&	-1.924	\\
26.5824	&	-1.150	&	26.5828	&	-1.609	&	26.5831	&	-1.950	&	27.5761	&	-1.105	&	27.5765	&	-1.556	&	27.5792	&	-1.946	\\
26.5845	&	-1.160	&	26.5849	&	-1.617	&	26.5853	&	-1.997	&	27.5784	&	-1.132	&	27.5789	&	-1.576	&	27.5815	&	-1.961	\\
26.5868	&	-1.182	&	26.5872	&	-1.627	&	26.5876	&	-2.002	&	27.5808	&	-1.152	&	27.5811	&	-1.591	&	27.5838	&	-1.965	\\
26.5916	&	-1.219	&	26.5919	&	-1.659	&	26.5923	&	-2.014	&	27.5830	&	-1.159	&	27.5834	&	-1.617	&	27.5863	&	-2.007	\\
26.5935	&	-1.211	&	26.5940	&	-1.661	&	26.5942	&	-1.983	&	27.5853	&	-1.194	&	27.5858	&	-1.636	&	27.5975	&	-2.034	\\
26.5954	&	-1.224	&	26.5958	&	-1.687	&	26.5960	&	-2.063	&	27.5967	&	-1.228	&	27.5971	&	-1.693	&	27.6002	&	-2.076	\\
26.5976	&	-1.250	&	26.5980	&	-1.694	&	26.5984	&	-2.041	&	27.5994	&	-1.250	&	27.5998	&	-1.720	&	27.6030	&	-2.061	\\
26.5998	&	-1.227	&	26.6002	&	-1.686	&	26.6006	&	-2.043	&	27.6021	&	-1.264	&	27.6025	&	-1.743	&	27.6058	&	-2.084	\\
26.6022	&	-1.250	&	26.6025	&	-1.710	&	26.6028	&	-2.049	&	27.6049	&	-1.270	&	27.6054	&	-1.725	&	27.6083	&	-2.110	\\
26.6044	&	-1.258	&	26.6049	&	-1.708	&	26.6052	&	-2.076	&	27.6077	&	-1.292	&	27.6081	&	-1.736	&	27.6111	&	-2.120	\\
26.6067	&	-1.266	&	26.6071	&	-1.708	&	26.6074	&	-2.093	&	27.6102	&	-1.302	&	27.6107	&	-1.733	&	27.6133	&	-2.095	\\
26.6089	&	-1.266	&	26.6092	&	-1.709	&	26.6096	&	-2.079	&	27.6125	&	-1.304	&	27.6130	&	-1.752	&	27.6169	&	-2.108	\\
26.6110	&	-1.282	&	26.6114	&	-1.725	&	26.6117	&	-2.094	&	27.6162	&	-1.302	&	27.6165	&	-1.768	&	27.6191	&	-2.115	\\
26.6132	&	-1.283	&	26.6137	&	-1.705	&	26.6140	&	-2.086	&	27.6183	&	-1.302	&	27.6187	&	-1.765	&	27.6214	&	-2.110	\\
26.6155	&	-1.279	&	26.6159	&	-1.714	&	26.6162	&	-2.057	&	27.6206	&	-1.290	&	27.6211	&	-1.742	&	27.6276	&	-2.085	\\
26.6177	&	-1.282	&	26.6182	&	-1.721	&	26.6186	&	-2.069	&	27.6268	&	-1.307	&	27.6272	&	-1.752	&	27.6299	&	-2.090	\\
26.6201	&	-1.284	&	26.6204	&	-1.718	&	26.6208	&	-2.085	&	27.6291	&	-1.280	&	27.6295	&	-1.731	&	27.6322	&	-2.069	\\
26.6223	&	-1.283	&	26.6228	&	-1.711	&	26.6231	&	-2.074	&	27.6314	&	-1.277	&	27.6318	&	-1.722	&	27.6349	&	-2.052	\\
26.6268	&	-1.272	&	26.6249	&	-1.730	&	26.6253	&	-2.066	&	27.6341	&	-1.256	&	27.6346	&	-1.735	&	27.6371	&	-2.082	\\
26.6291	&	-1.253	&	26.6273	&	-1.719	&	26.6276	&	-2.046	&	27.6363	&	-1.264	&	27.6367	&	-1.703	&	27.6393	&	-2.085	\\
26.6313	&	-1.246	&	26.6295	&	-1.718	&	26.6321	&	-2.066	&	27.6408	&	-1.232	&	27.6389	&	-1.698	&	27.6415	&	-2.032	\\
26.6336	&	-1.261	&	26.6317	&	-1.704	&	26.6391	&	-2.044	&	27.6430	&	-1.261	&	27.6413	&	-1.682	&	27.6438	&	-1.990	\\
26.6382	&	-1.229	&	26.6340	&	-1.686	&	26.6437	&	-1.967	&	27.6666	&	-1.052	&	27.6435	&	-1.656	&	27.6484	&	-2.022	\\
26.6406	&	-1.192	&	26.6387	&	-1.667	&	26.6485	&	-1.925	&	27.6689	&	-0.987	&	27.6480	&	-1.720	&	27.6521	&	-1.977	\\
26.6428	&	-1.188	&	26.6410	&	-1.677	&	26.6507	&	-1.957	&		&		&		&		&		&		\\
\hline
\end{tabular}
\end{center}
\end{table}

\begin{table}
\scriptsize
\caption{The depths of eclipse (differential magnitude) in each band.}
\begin{center}
\begin{tabular}{lcc}\hline\hline
Filters&Nov 2016 (pri, sec)&Mar 2017 (pri, sec)\\
\hline
$\Delta B$ ($\pm$ 0.004)& 0.280 0.127 & 0.288 0.089\\
$\Delta V$ ($\pm$ 0.002)& -0.541 -0.682 & -0.536 -0.681\\
$\Delta R_C$($\pm$ 0.002)&-1.052 -1.158&-1.033 -1.151\\
$\Delta I_C$($\pm$ 0.003)&-1.402 -1.491&-1.389 -1.487\\
\hline
\end{tabular}
\end{center}
\end{table}

\section{Orbital period investigations}

\begin{figure}
\begin{center}
\includegraphics[angle=0,scale=0.5]{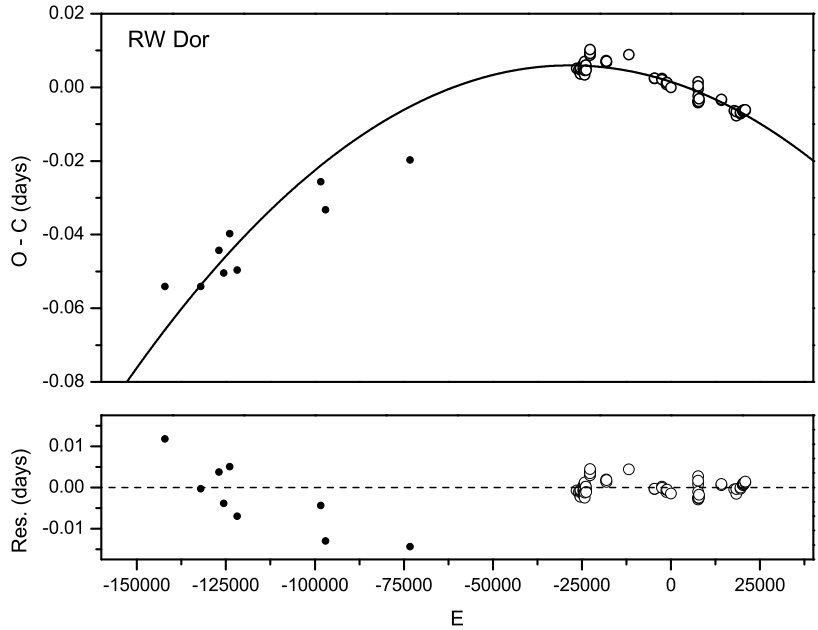}
\caption{The $(O-C)$ curve. The dots before E = -50000 refer to photographic data (pg) and the open circles after E = -50000 refer to photoelectric (pe) and CCD data. The solid line in the upper panel was computed by using the quadratic term in Eq. (2) and this downward parabolic curve suggests a long-term period decrease. The residuals from Eq. (2) are plotted in the lower panel.}
\end{center}
\end{figure}

\begin{figure}
\begin{center}
\includegraphics[angle=0,scale=0.5]{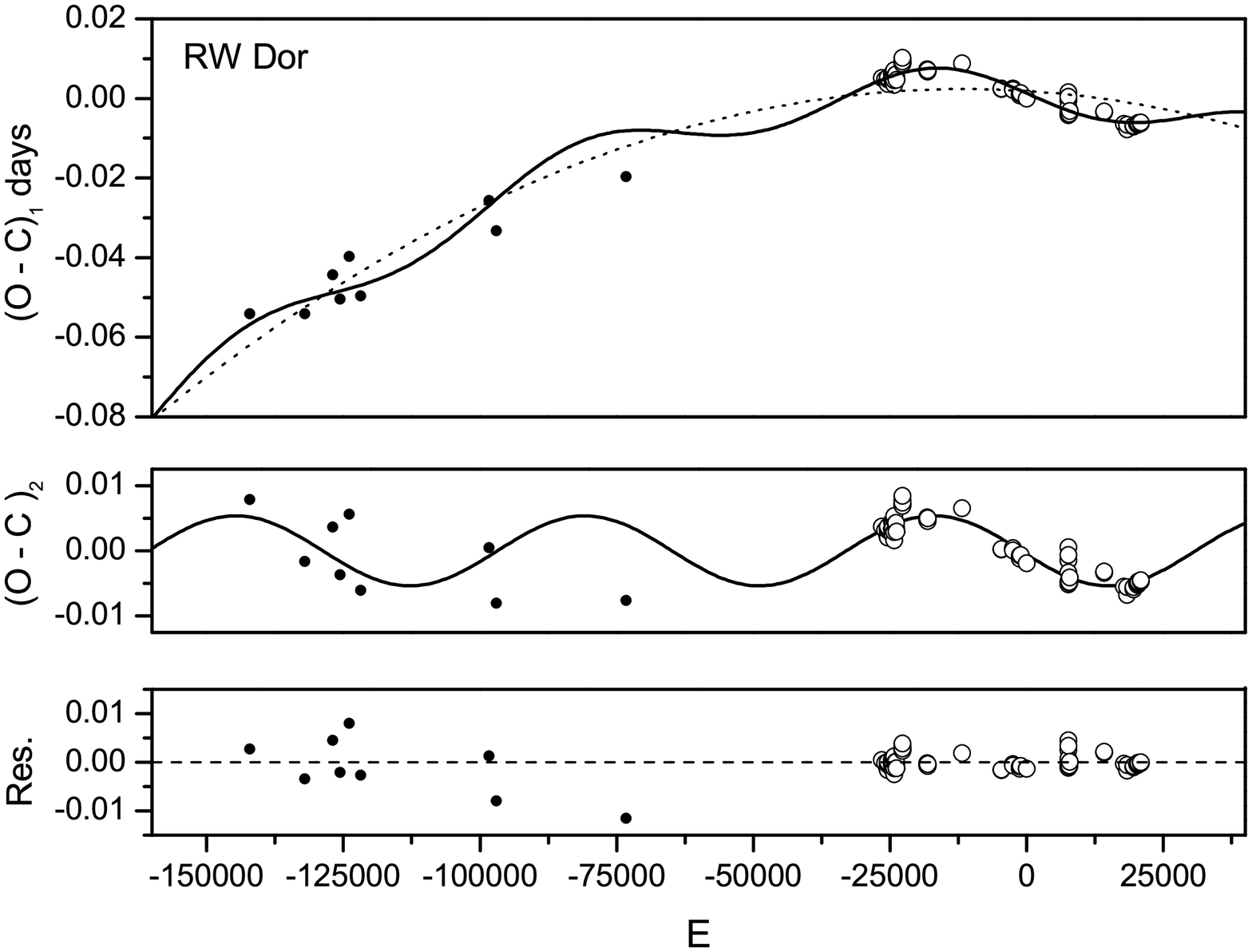}
\caption{The $(O-C)_1$ diagram was constructed by using the ephemeris in Eq. (1). The solid line in upper panel refers to a combination of a long-term period decrease and a small-amplitude cyclic variation, while the dashed line represents the long-term decrease of the orbital period.}
\end{center}
\end{figure}

The orbital period changes were analysed by using the $O-C$ method. All available eclipse times from the literature are collected and investigated together with new CCD photometric times from present observations listed in Table 5. The $O-C$ values of those eclipse times were computed by using the linear ephemeris given by Kreiner (2004);
\begin{equation}
Min.I (HJD) = 2451869.076+0^{\textrm{d}}.2854633E.
\end{equation}
To analyse the $O-C$ changes, we firstly use a quadratic ephemeris to fit the $O-C$ curve with weights 1 to photographic data (pg) and weights 10 to photoelectric (pe) or CCD data. The result is plotted as the solid line in the upper panel of Fig. 2. A least-squares solution leads to the following quadratic ephemeris:
\begin{equation}
\begin{array}{ll}
Min.I (HJD) = 2451869.07751(\pm 0.00004) + 0.285462985(\pm 0.000000002)E
\\            -[55.5(\pm 0.2)\times 10^{-13}]E^2.
\end{array}
\end{equation}
From the quadratic term in Eq. (2), the orbital period is decreasing and the change rate can be determined as $\mathrm{d}P/\mathrm{d}t = -14.19(\pm 0.05)\times10^{-9}$ d $\textrm{yr}^{-1}$. The residuals from Eq. (2) are plotted in the lower panel of Fig. 2. As shown in in Fig. 2, we found that just only parabolic curve may not fit well, a cyclic variation seems to exist which can be seen clearly from the residuals in lower panel that are showing a large systematic scatter. Therefore, to get a better fit for the trend of $O-C$, a sinusoidal term is added to a quadratic ephemeris. By using a least-squares method, the ephemeris is re-determined as,
\begin{equation}
\begin{array}{lll}
Min.I(HJD) =  2451869.07786(\pm 0.00004)+ 0.285463214(\pm 0.000000007)E
\\            -[37.6(\pm 0.5)\times 10^{-13}]E^2
\\            +0.0054(\pm 0.0001) \times \sin[0^{\circ}.00564E+185^{\circ}.824(\pm 0^{\circ}.684)].
\end{array}
\end{equation}

As shown in Fig. 3, Eq. (3) can give a good description to the general trend of the $O-C$ curve. After the downward parabolic change is subtracted, the cyclic oscillation is displayed in the middle panel. From this ephemeris, the sinusoidal term suggests that the semi-amplitude of cyclic variation is about 0.0054 days, while it has a long period of 49.92 years. The quadratic term also reveals a continuous period decrease at a rate of $\mathrm{d}P/\mathrm{d}t = -9.61(\pm 0.13) \times10^{-9}$ d $\textrm{yr}^{-1}$ that is smaller than that derived from Eq. (2). The residuals from Eq. (3) are also plotted in the bottom panel of Fig. 3. Its scatters are smaller than those in Fig. 2 {\bf that can be justified by the sum of weighted squares deviation as 0.000851 for Eq. (2) without cyclical term and as 0.000462 for Eq. (3) with cyclical term, respectively.}

\begin{table}
\scriptsize
\caption{Times of light minimum for RW Dor.}
\begin{center}
\begin{tabular}{llcccllccc}\hline\hline
HJD &Error &Method &Min. &Ref. &HJD &Error &Method &Min. &Ref.\\
(2400000+) &(days) & &(type) & &(2400000+) &(days) & &(type) &\\
\hline
11298.835   &        & pg & II & (1) &46681.7865  & 0.001 & pe & II & (5)\\
14168.883   &        & pg & II & (1) &46690.7785  & 0.0001& pe &  I & (6)\\
15621.901   &        & pg & II & (1) &46695.7745  & 0.0001& pe & II & (6)\\
16013.836   &        & pg & II & (1) &48500.0470  &       &    &  I& (10)\\
16489.714   &        & pg & II & (1) &50559.9437  & 0.0004&ccd &  I & (7)\\
17075.903   &        & pg &  I & (1) &50560.0865  & 0.0003&ccd & II & (7)\\
23784.600   &        & pg &  I & (1) &51158.5603  & 0.0002&ccd &  I & (7)\\
24172.537   &        & pg &  I & (1) &51158.7027  & 0.0002&ccd & II & (7)\\
30938.602   &        & pg &  I & (10)&51505.6820  & 0.0002&ccd &  I & (7)\\
44313.581   & 0.001  & pe & II & (2) &51505.8252  & 0.0002&ccd & II & (7)\\
44464.876   & 0.001  & pe & II & (2) &51548.5020  &       &ccd &  I &(10)\\
44581.7728  & 0.0008 & pe &  I & (2) &51869.0760  &       &ccd &  I& (10)\\
44608.6063  & 0.0005 & pe &  I & (2) &54036.5947  &       &ccd &  I& (10)\\
44608.7488  & 0.0006 & pe & II & (2) &54036.7411  &       &ccd& II & (10)\\
44609.6063  & 0.0008 & pe & II & (2) &54037.5995  &       &ccd& II & (10)\\
44609.7493  & 0.0005 & pe &  I & (2) &54037.7384  &       &ccd&  I & (10)\\
44610.7487  & 0.0009 & pe & II & (2) &54041.7335  &       &ccd&  I & (10)\\
44825.8462  & 0.0008 & pe &  I & (3) &54049.5878  &       &ccd& II & (10)\\
44826.8442  & 0.0004 & pe & II & (3) &54059.7177  &       &ccd&  I & (10)\\
44873.8038  & 0.0006 & pe &  I & (3) &54087.9783  & 0.0001& pe &  I & (8)\\
44874.6594  & 0.0003 & pe &  I & (3) &54091.1189  & 0.0002& pe &  I & (8)\\
44874.8010  & 0.0004 & pe & II & (3) &54095.1150  & 0.0001& pe &  I & (8)\\
44958.5851  & 0.0004 & pe &  I & (3) &54107.6761  &       & ccd&  I & (10)\\
44961.5843  & 0.0006 & pe & II & (3) &55904.66718 &0.00005&ccd &  I & (9)\\
44961.7239  & 0.0006 & pe &  I & (3) &55906.80840 &0.00005&ccd & II & (9)\\
44962.5815  & 0.0007 & pe &  I & (3) &56950.7446  &       & ccd& II & (10)\\
44962.7267  & 0.0005 & pe & II & (3) &57112.6009  & 0.0001 &ccd & II & (9)\\
45021.6738  & 0.0003 & pe &  I & (3) &57118.5968  & 0.0001 &ccd & II & (9)\\
45049.5058  & 0.0006 & pe & II & (3) &57446.7363  & 0.0001 &ccd &  I & (9)\\
45049.6486  & 0.0008 & pe &  I & (3) &57447.5930  & 0.0001 &ccd &  I & (9)\\
45050.6484  & 0.0003 & pe & II & (3) &57644.8485  & 0.0002 &ccd &  I & (9)\\
45076.4815  & 0.0004 & pe &  I & (3) &57645.8477  & 0.0001 &ccd & II & (9)\\
45370.6556  & 0.0001 & pe & II & (4) &57661.8334  & 0.0002 &ccd & II & (9)\\
45370.6558  & 0.0003 & pe & II & (4) &57686.6689  & 0.0003 &ccd & II & (9)\\
45370.6564  & 0.0004 & pe & II & (4) &57686.8119  & 0.0002 &ccd &  I & (9)\\
45376.6502  & 0.0001 & pe & II & (4) &57720.6390  & 0.0002 &ccd & II & (9)\\
45376.6507  & 0.0002 & pe & II & (4) &57721.6383  & 0.0002 &ccd &  I & (9)\\
45376.6517  & 0.0002 & pe & II & (4) &57826.5461  & 0.0002 &ccd & II & (9)\\
46680.7878  & 0.001  & pe &  I & (5) &57827.5454  & 0.0002 &ccd &  I & (9)\\
\hline
\end{tabular}
\end{center}
{\tiny Notes.} \tiny (1) Hertzsprung 1928, (2) Marton \& Grieco 1981, (3) Grieco \& Marton 1983, (4) Russo et al. 1984, (5) Marton et al. 1989, (6) Kaluzny \& Caillault 1989, (7) Ogloza \& Zakrzewski 2004, (8) Marino et al. 2007,(9) the present authors, (10) http://var.astro.cz/ocgate.
\end{table}

However, it cannot conclude that the result from Fig. 3 is better and reliable than Fig. 2. Because of small number of the $O-C$ data that are not cover (the gap between 1943 and 1980 or about 36.64 years) the whole cycles. Furthermore, the period of the third body from Fig. 3 is 49.92 years, while the total time span of available eclipse times only 127.48 years. Besides, the first nine minima (pre-1944) are photographic (pg) which listed in Table 3 with only three decimal places, thus they are ten times smaller weights when compared to the photoelectric (pe) and CCD data. Consequently, the effective time span of higher-quality minima (pe and CCD data) is only 37 years, less than the third body orbit period. Moreover, the point density (9 minima over a time interval of 53.81 years) for the first group is much lower than that of the second group (69 minima over a time interval of 37 years). For these reasons, the evidence for a sinusoidal term in Fig. 3 is very weak. The result from Eq. 2 may be possible for explanation of period change in RW Dor system. However, to check the period changes proposed here, more eclipse times are still required in the future to confirm the period variations.

\section{Photometric solutions}

As shown in Fig. 1, the two sets of light curves obtained in November 2016 and March 2017 overlap nearly. This suggests that the light curves are stable within the error at that time interval and no light curve variations {\bf from O'Connell effect or spot activity cycles} are found. However, the light curves of some close binaries are asymmetric that could be explained by spot activity on one or both components (e.g., Qian et al. 2017a). For RW Dor, the light curves are quite symmetric indicating that they are {\bf very useful to determine the photometric solutions with high accuracy and it does not need the spot model which is more complicate solution}. We model the two sets of light curves separately, i.e., set-1 refers to the one observed on 28-29 November 2016 and set-2 refers to the other obtained on 14-15 March 2017. The two data sets in $B$, $V$, $R_{c}$ and $I_{c}$ bands were analysed separately by using the Wilson \& Devinney (W-D) code (Wilson \& Devinney 1971; Wilson 1990, 1994, 2012; van Hamme \& Wilson 2007).

\begin{figure}
\begin{center}
\includegraphics[angle=0,scale=0.3]{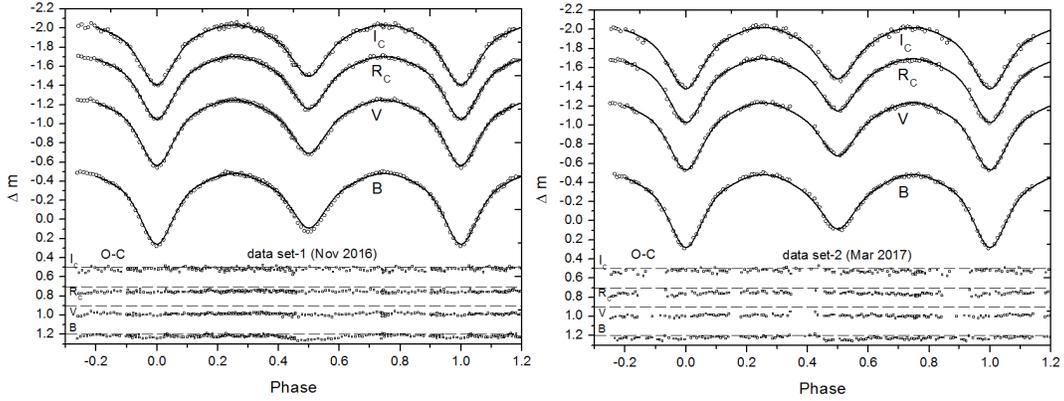}
\caption{Theoretical light curves (solid lines) computed using the W-D method compared to the observed light-curves for data set-1 (left panel) and for data set-2 (right panel) without third light and spot. {\bf All theoretical light curves in BVRI fit well with the observed light curves, this means that the physical parameters obtained from light curves modeling are correct and reliable}, except $B$-band light curves in data set-1 which is not fit well at the secondary minimum.}
\end{center}
\end{figure}

\begin{figure}
\begin{center}
\includegraphics[angle=0,scale=0.5]{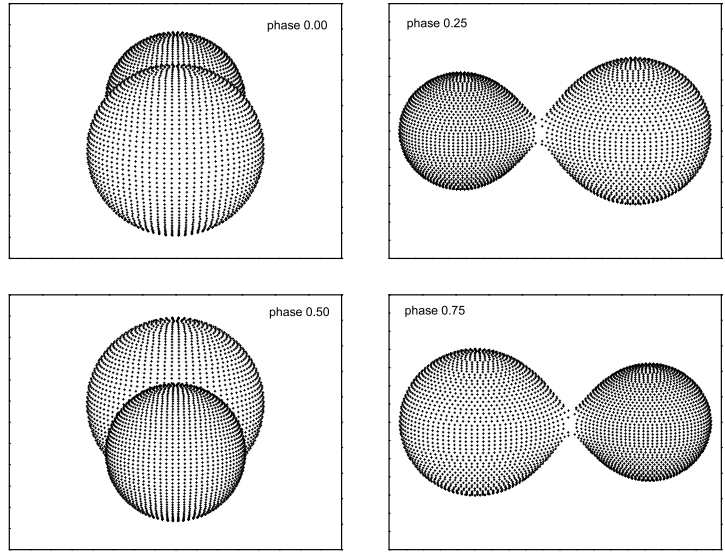}
\caption{Geometrical structures at phases of 0.00, 0.25, 0.50 and 0.75}
\end{center}
\end{figure}

Hilditch et al. (1992) classified the spectral type of the binary as K1 V following Marton et al. (1989). The Tycho-2 mean color index $B-V$=0.66 (Hog et al. 2000) corresponds to a spectral type of G4/5\,V, while the color index in SIMBAD database reveals a $B-V$ = 0.69. For our solutions, the effective temperature of the primary star ($T_1$) was assumed at 5560 K according to its spectral type (Cox 2000). We assume that the convective envelopes are already developed. Thus, the bolometric albedos for star 1 and 2 were taken as $A_{1} = A_{2}$ = 0.5 and the values of the gravity-darkening coefficients $g_{1} = g_{2}$ = 0.32 were used. The monochromatic and bolometric limb-darkening coefficients were taken with logarithmic law from van Hamme's table (van Hamme 1993). All fixed parameters are listed in Tables 4. The adjustable parameters are the inclination ($i$), the mass ratio ($q$), the temperature of star 2 ($T_{2}$), the monochromatic luminosity of star 1 ($L_{1B}$, $L_{1V}$, $L_{1R_{c}}$ and $L_{1I_{c}}$), the dimensionless potential of star 1 ($\Omega_{1}=\Omega_{2}$) in mode 3 for contact configuration.

\begin{table}
\scriptsize
\caption{Photometric solutions for data set-1 and set-2 when $T_{1}$=5560 K.}
\begin{center}
\begin{tabular}{lcccc}\hline\hline
Parameters  & set-1 & set-2\\
\hline
$T_1(K)$  & 5560{\bf(assumed)} & fixed\\
$g_1=g_2$ & 0.32{\bf(assumed)} & fixed \\
$A_1=A_2$ & 0.50{\bf(assumed)} & fixed \\
$q(M_{2}/M_{1})$ &1.587{\bf(assumed)}& fixed \\
$T_2$ (K) & 5287($\pm10$) &5238($\pm13$) \\
$T_1 - T_2$ (K) & 273{\bf($\pm5$)} & 322{\bf($\pm6$)} \\
$T_2/T_1$ & 0.951{\bf($\pm0.002$)} & 0.942{\bf($\pm0.002$)} \\
$i(^o)$&77.2($\pm0.1$) &76.9($\pm0.2$) \\
$\Omega_{in}$  & 3.091 & 3.091 \\
$\Omega_{out}$ & 2.732 & 2.732 \\
$\Omega_1=\Omega_2$ &4.64($\pm0.04$) &4.62($\pm0.03$)\\
$L_1/(L_1+L_2)$($B$)&0.475($\pm0.006$) &0.491($\pm0.006$) \\
$L_1/(L_1+L_2)$($V$)&0.455($\pm0.005$) &0.467($\pm0.005$) \\
$L_1/(L_1+L_2)(R_{c})$ &0.444($\pm0.005$)&0.455($\pm0.005$)\\
$L_1/(L_1+L_2)(I_{c})$ &0.437($\pm0.005$)&0.446($\pm0.004$)\\
$r_1(pole)$&0.3222($\pm0.0013$)&0.3242($\pm0.0014$)\\
$r_1(side)$&0.3378($\pm0.0014$)&0.3402($\pm0.0015$)\\
$r_1(back)$&0.3740($\pm0.0015$)&0.3777($\pm0.0018$)\\
$r_2(pole)$&0.4057($\pm0.0053$)&0.4069($\pm0.0048$)\\
$r_2(side)$&0.4309($\pm0.0069$)&0.4326($\pm0.0063$)\\
$r_2(back)$&0.4635($\pm0.0102$)&0.4658($\pm0.0093$)\\
$f$ &11.5$\%$($\pm6.7\%$) & 15.0$\%$($\pm6.1\%$)\\
$\Sigma{W_{i}(O-C)_{i}^2}$ &0.01641& 0.01866 \\
\hline
\end{tabular}
\end{center}
\end{table}

For precise mass ratio determination, Hilditch et al. (1992) had published results of radial-velocity measurements and gave a mass ratio of $q_{sp}=0.68\pm0.03$. Later, Duerbeck \& Rucinski (2007) obtained $q_{sp}=0.63\pm0.03$. However, we used a $q$-search method first with our photometric data to determine the photometric mass ratio ($q_{ph}$) and then set mass ratio as an adjustable parameter to get a better fit for data set-1 and set-2. We obtained the initial mass ratio at $q$ = 1.6 {\bf(q $>$ 1.0 for W-subtype)} and then the differential correction was performed until the final solutions were derived at the lowest sum of the weighted square deviations $\Sigma (W(O-C))^2$, hereafter $\Sigma$. The result is about $q$ {\bf (W-subtype) = 1.62$\pm0.02$ or $q$ = 0.615 in general meaning} (which is close to the spectroscopic mass ratio of 0.63 that derived by Duerbeck \& Rucinski, 2007) for both data sets {\bf when the more massive component of W-subtype system is the cooler}. However, it is obvious that both sets of light curves show partial eclipses ($i<$ 85 deg) that theirs photometric mass ratios obtained by $q$-search may not be accurate as discussed by Terrell \& Wilson (2005). Therefore, we use the spectroscopic mass ratio 0.63 or {\bf $q$ (W-subtype) = 1.587} for our fixed mass ratio in the modeling process. The light-curve modeling results with optimum parameters are listed in Table 6. The corresponding theoretical light curves (solid lines) were compared to the observed light curves as shown in Fig. 4 for data set-1 (left panel) and for data set-2 (right panel). Additionally, Fig. 3 shows a cyclic variation that may be caused by light-travel time effect via the presence of a third body. Therefore, we added the third light ($l_{3}$) as an adjustable parameter in order to check the luminosity contribution of such a third companion, but the results show the negative values for both data sets. This may suggest that if the third body really exists, it will be a very cool stellar companion. On the other hand, it may have no any companion orbiting the eclipsing pair. The presence of the third body will be discussed again in the next section. The geometrical structures of RW Dor based on the modeling are displayed in Fig. 5.

\begin{table}
\scriptsize
\caption{Parameters of the tertiary component for RW Dor.}
\begin{center}
\begin{tabular}{lrcl}\hline\hline
 Parameters & Value & Error & Units\\
\hline
$P_{3}$ & 49.9207 & 0.0003 & yrs\\
$A_{3}$ & 0.0054 & {\bf 0.0001} & days\\
$a'_{12} \sin{i'}$ & 0.929 & 0.026 & AU\\
$f(m)$ & 0.00032 & 0.00003 & $M_{\odot}$\\
$e_3$ & 0.0 & assumed & -\\
$M_3$ ($i'=90^{\circ}$) & 0.087 & 0.002 & $M_{\odot}$\\
$a_3$ ($i'=90^{\circ}$) & 14.33 & 0.58 & AU\\
\hline
\end{tabular}
\end{center}
\end{table}

\section{Discussions and conclusions}

The two sets of complete multi-color light curves in $BV(RI)_c$ bands were obtained by using the PROMPT-8 robotic telescope at CTIO in Chile from March 2015 to March 2017. The other data in 2011 were obtained by using the 2.15-m telescope at CASLEO, San Juan in Argentina. We compare our results to the light curves published by Deb \& Singh (2011), no O'Connell effect and light curve variations were found. The photometric solutions indicate that RW Dor is a W-subtype, shallow contact binary with a degree of contact more than $10\%$ and a high mass ratio {\bf $q$ (W-subtype) $\sim$ 1.587 or $q$ = 0.63 in general meaning}, which indicates that the hotter component is the less massive one. The absolute dimensions of RW Dor are derived by using our photometric elements together with spectroscopic one by Duerbeck \& Rucinski (2007), the results are: $M_{1}=0.52M_{\odot}$, $M_{2}=0.82M_{\odot}$, a=2.03$R_{\odot}$, $R_{1}=0.703R_{\odot}$, $R_{2}=0.881R_{\odot}$, $L_{1}=0.423L_{\odot}$ and $L_{2}=0.534L_{\odot}$. These parameters are close to those recently derived by Deb \& Singh (2011).

The downward parabolic curve in the $O-C$ diagram shows the orbital period decrease at a rate of $\mathrm{d}P/\mathrm{d}t = -14.19\times10^{-9}$ d $\textrm{yr}^{-1}$ without sinusoidal term and $\mathrm{d}P/\mathrm{d}t = -9.61\times10^{-9}$ d $\textrm{yr}^{-1}$ with {\bf a cyclical term}. The type of variations, i.e., a long-term decrease combined with a cyclic change, is commonly found in W UMa-type stars, for examples, V417 Aql (Qian 2003), V1139 Cas (Li et al., 2015a), MR Com (Qian et al. 2013a), BX Peg (Li et al. 2015b), V524 Mon (He et al. 2012), and V1073 Cyg (Tian et al. 2018). Some W-type contact binaries whose properties are similar to RW Dor are listed in Table 8, most of them are shallow contact binaries with decreasing period. The long-term period decrease can be explained either by the mass transfer from the more massive component to the less massive one or by the angular momentum loss (AML) via magnetic braking, or by the combination of both processes. If the long-term period decrease is due to conservative mass transfer, the mass transfer rate can be determined with the following equation {\bf(Kwee 1958)},
\begin{equation}
\frac{\dot{P}}{P} = - 3 \dot{M_2}(\frac{1}{M_1}-\frac{1}{M_2})
\end{equation}
The mass transfer rate is $\mathrm{d}M/\mathrm{d}t = 23.55\times10^{-9}$ $M_{\odot} \textrm{yr}^{-1}$ {\bf without cyclical term} and $\mathrm{d}M/\mathrm{d}t = 15.95\times10^{-9}$ $M_{\odot} \textrm{yr}^{-1}$ {\bf with cyclical term}. The timescale of mass transfer can be computed as $M_2/\dot{M} \sim 3.48\times10^{7} \textrm{yr}$ (or 35 Myr) for Eq. 2 and $M_2/\dot{M}\sim 5.14\times10^{7} \textrm{yr}$ (or 51 Myr) for Eq. 3. While the time scale of period decrease $P/\dot{P} \sim 2.01\times10^{7} \textrm{yr}$ (or 20 Myr) {\bf without sinusoidal term} and $P/\dot{P} \sim 2.97\times10^{7} \textrm{yr}$ (or 30 Myr) {\bf with sinusoidal term}. The thermal timescale of the more massive component is $46.43\times10^{6} \textrm{yr}$ (or 46 Myr). Those timescales reveal that RW Dor is presently undergoing a slow mass transfer at the beginning stages of contact evolution with high mass ratio, shallow contact configuration and long-term orbital period decrease. In this way, the contact degree of the system will become higher and the system will evolve into a deeper contact binary. Another plausible explanation for long-term period decrease is AML via magnetic stellar wind and it can be determined by the following equation given by Bradstreet \& Guinan (1994),
\begin{equation}
\dot{P} \approx -1.1\times10^{-8}q^{-1}(1+q)^2(M_1 + M_2)^{-5/3}k^2
         \times(M_{1}R^{4}_{1} + M_{2}R^{4}_{2})P^{-7/3},
\end{equation}
where $k^2$ is the gyration constant ranging from 0.07 to 0.15 for solar type stars. By adopting a value of $k^2$ = 0.1 (Bradstreet \& Guinan 1994), the rate of orbital period decrease due to AML can be computed as $\mathrm{d}P/\mathrm{d}t = -33.2\times10^{-9}$ d $\textrm{yr}^{-1}$, in this case the timescale of period decrease is $P/\dot{P} \sim 8.598\times10^{6} \textrm{yr}$ (or 8.6 Myr) which is about two times shorter than the timescale from Eq. 2 and three times from Eq. 3. This means that the conservative mass transfer may not satisfy to explain the secular period decrease or the mass transfer may be dynamical (Qian \& Zhu 2002). To explain this, Qian (2001a) proposed that the rate of AML is changed depending on the depth of overcontact. When the period decrease, the separation between the components becomes closer and the depth of contact increases, causing common convective envelope (CCE) to become deeper and increase mixing in the CCE which may result to AML to be lower rate (Vilhu 1981; Smith 1984). If AML value is larger than the critical value of Rahunen (1981) and causes orbital period to decrease, the evolution of RW Dor will be on the AML-controlled stage. Based on period studies by Qian (2001a), the evolution of RW Dor may be the combination of the thermal relaxation oscillation (TRO) and the AML changes through the variable depth of overcontact, e.g. V417 Aql (Qian 2003). In addition, the study by Marton et al. (1989) has shown that there is a hot spot on the cooler component located near the neck of the system, suggesting a secondary-to-primary mass transfer which has a good agreement with the long-term period decrease. This indicates that RW Dor is in the transition phase to W UMa and is at the beginning of the contact phase similar to VW Boot (Qian \& Zhu 2002). If the orbital period decrease is caused by losing angular momentum through magnetic braking, this is in agreement with the conclusion derived by Qian et al. (2017b, 2018) that some EW-type contact binaries were formed from short-period EA-type systems via case A mass transfer (Bradstreet \& Guinan 1994; Vilhu 1982). Based on spectroscopic observations with LAMOST (Qian et al. 2017b, 2018) reveal that short period EW binaries (P $<$ 0.4 d) have low metallicities, suggesting that EW-type binaries are old stellar populations and may be older than their long period cousins. This means that they have a longer pre-contact phase (Qian et al. 2017b). Moreover, the evolution study of low-mass contact binaries (LMCB) by Stepien \& Gazeas (2012) indicates that the systems with low total mass (M $<$ 1.4 $M_{\odot}$) and short orbital period (P $<$ 0.3 d) have a long pre-contact phase that lasts for 8-9 Gyrs, while the contact phase takes only about 0.8 Gyr with low mass transfer rate. The situation of RW Dor has good agreements with the conclusions proposed by Stepien \& Gazeas (2012) and Qian et al. (2017b, 2018) that the shallow-contact binary with short period (P $\sim$ 0.285 d), low total mass (M $\sim$ 1.34 $M_{\odot}$) and long-term period decrease is a newly formed contact binary, which is similar to V524 Mon (He et al. 2012), MR Com (Qian et al. 2013a), BI Vul (Qian et al. 2013b) and CSTAR 038663 (Qian et al. 2014a), with the beginning state of the contact phase or recently evolved into contact configuration after it spent a long time in pre-contact phase. In addition, the absolute parameters of RW Dor are quite close to V336 TrA (Kriwattanawong et al 2018, see Fig. 4), suggesting that the more massive component (the secondary) of RW Dor will locate near the ZAMS (the zero age main sequence), while the location of the less massive one should be close to the TAMS (the terminal age main sequence). This means that the less massive one has evolved to reach the TAMS, whereas the more massive one has not evolved. It may be due to the mass transfer between the components.

\begin{table}
\scriptsize
\caption{Parameters of marginal contact binaries (W-type systems).}
\begin{center}
\begin{tabular}{lllcrcccccccc}\hline\hline
Star &Sp. &Period & $1/q$ & $f$ &$\mathrm{d}P/\mathrm{d}t$ &Cyclic &$l_3$ & $M_1$ & $M_2$ & $M_3$ & activities&Ref.\\
 & & (days)& &($\%$)&($\times 10^-{8}$ d/y)& & &($M_{\odot}$)&($M_{\odot}$)&($M_{\odot}$)&\\
\hline
KIC 9532219&G9&0.1981&0.833&-&-52.7&yes&76$\%$&-&-&0.09&spot&(1)\\
CC Com&K4/5&0.2207&0.527&16.7&-20.0&yes&-&0.37&0.71&0.066&-&(2)\\
BI Vul&K3&0.2518&0.964&8.7&-9.5&yes&no&0.72&0.75&0.30&spot&(3)\\
V336  TrA&K1&0.2668&0.716&15.7&-&-&-&0.65&0.91&-&spot&(4)\\
CSTAR 038663&K3&0.2671&0.893&10.6&-&yes&$<1\%$&0.72&0.81&0.63/2.02& spot/flares &(5)\\
BM UMa&K&0.2712&0.540&17.0&-7.49&yes&-&0.50&0.92&0.25&-&(6)\\
BX Peg &G5&0.2804 &0.372&23.1 &-9.84&yes&no&0.38&1.02&0.22&spot&(7)\\
GSC 2765-0348&G4&0.2835&0.313&34.0&-&-&-&-&-&-&2 spots&(8)\\
V524 Mon &G5&0.2836&0.476&7.7 &-.015 &yes&no&0.50&1.10&0.26&no&(9)\\
LO Com&K0&0.2864&0.404&3.2&-11.8&yes&no&0.32&0.79&-&no&(10)\\
GSC 3526-01995&K2&0.2922&0.351&18.2&-&yes&no&0.28&0.80&0.57&spot&(11)\\
IK Boo&G2&0.3031&0.873&2.2&-21.7&yes&-&0.86&0.99&0.21&spot&(12)\\
V2284 Cyg&G7&0.3069&0.345&39.2 &-29.7 &yes&no&0.30&0.86&0.036&no&(13)\\
TY Boo&G3&0.3171&0.466&12.0&-2.99&yes&-&0.53&1.14&0.49&-&(14)\\
V1007 Cas&K0&0.3320&0.297&8.1&-17.8&-&1.1$\%$&0.34&1.14&-&spot&(15)\\
V781 Tau&G0&0.3449&0.453&21.6&-6.01&yes&no&0.71&1.57&0.16&spot&(16)\\
V396 Mon&F8&0.3963&0.392&18.9&-8.57&yes&no&0.36&0.92&0.31&-&(17)\\
PP Lac &G6&0.4012&0.435&23.9& - &yes&no&0.51&1.18&0.21&no&(18)\\
MR Com &F5&0.4127&0.256&10.0&-53.0& yes&$<1\%$&0.36&1.40&0.18&no&(19)\\
RW Dor &G4/5&0.2854&0.630&$>$10&-1.42&unclear&no&0.52&0.82&unclear&no&(20)\\
\hline
\end{tabular}
\end{center}
{\tiny Notes.} \tiny (1) Lee et al 2016, (2) Yang et al. 2009, (3) Qian et al. 2013b, (4) Kriwattanawong et al. 2018, (5) Qian et al. 2014a, (6) Yang et al 2009, (7) Lee et al. 2004 \& 2009, (8) Samec et al. 2012,(9) He et al. 2012, (10) Zhang et al. 2016, (11) Liao et al. 2012, (12) Kriwattanawong et al. 2017, (13) Wang et al. 2017, (14) Yang et al. 2007, (15) Li et al. 2018, (16) Li et al. 2016, (17) Liu et al. 2011, (18) Qian et al. 2005, (19) Qian et al. 2013a, (20) present study.
\end{table}

In addition, a cyclic oscillation superimposed on a secular term is {\bf often} found in W UMa type binaries (see Qian 2001b, Qian 2002). {\bf Furthermore, a study of the light travel time effect (LTTE) in short period eclipsing binaries by Li et al. (2018) indicates that the frequency of third bodies found in contact binaries with period shorter than 0.3 days reaches a value of 0.65 (65 \%) in their samples of 542 eclipsing binaries.} Therefore, it is possible that there may exist a periodic variation superimposed on a secular period decrease in the $O-C$ curve of RW Dor, even weak evidence as explained in section 3. If assuming that the periodic change in $O-C$ curve exists, by analysis of Eq. 3, the sinusoidal term reveals a cyclic change with a semi-amplitude of 0.0054 days and a period of 49.92 years. The periodic variations in W UMa binary stars are usually explained by the two ideas, one is the Applegate mechanism (Applegate 1992) via magnetic activity cycles at one or both components. The second idea is the light-travel time effect (Liao \& Qian 2010; Han et al. 2016) via the presence of a third body.

The Applegate mechanism suggests that the cyclic change is caused by magnetic activity-driven variations in the quadrupole moment of solar-type components. Because RW Dor consists of G4/5 V spectral type stars, it should show strong magnetic activity. If this is in the case, the quadrupole moment of the binary star can be determined from the equations given by Rovithis-Livaniou et al. (2000) and Lanza \& Rodono (2002),
\begin{equation}
\Delta{P} = \sqrt{[1-\cos(2\pi P/P_3)]}\times A_3
\end{equation}
and
\begin{equation}
\frac{\Delta{P}}{P} = - 9 \frac{\Delta{Q}}{Ma^2},
\end{equation}
where $A_3$ is the amplitude of the sinusoidal term in Eq. (3), $P_3$ is the magnetic activity period, $R$ is the radius of the active star and $a$ is the separation. The result is $\Delta{P}/P$ $\sim$ $1.317\times 10^{-6}$ and the quadruple moment for primary star $\Delta{Q}_1$ = $3.07\times10^{48}$ \textrm{g} $\textrm{cm}^2$ and for secondary star $\Delta{Q}_2$ = $4.84\times10^{48}$ \textrm{g} $\textrm{cm}^2$. These values for both components are lower than the typical values for active contact binary which ranging from $10^{51}$ to $10^{52}$ \textrm{g} $\textrm{cm}^2$. Thus, the cyclic oscillation in Fig. 3 cannot be interpreted by the result of Applegate mechanism. Furthermore, no spot activity cycles or light curve variations were found during the observations. The light curves may be stable for years (see Fig. 1) comparing to the previous publications, which suggests that there is very weak magnetic activity cycle at that time. Therefore, this period modulation may not be caused by magnetic activity cycle that happens normally in single solar-type stars. There are many contact systems that periodic variations cannot be explained by the Applegate mechanism as discussed by Qian et al. (2013b), but the most probable reason causing the cyclic changes for those binaries is the light-travel time effect due to the perturbations from a third body (Irwin 1952, Borkovits \& Hegedues 1996, Liao \& Qian 2010).

Therefore, the plausible idea for the cyclic period change is the light-travel time effect via the presence of a third body. By assuming that the tertiary component is moving in a circular orbit, the value of $a'_{12} \sin{i'}$ is computed as 0.929\,au by using the relation $a'_{12} \sin{i'}$ = $A_3 \times c$, where $A_{3}$ is the semi-amplitude of the $O-C$ oscillation, $c$ is the speed of light and ${i'}$ is the orbital inclination of the third component. Thus the mass function, the masses and the orbital radii of the third component in different inclination values can be determined with the following equation:
\begin{equation}
f(m) = \frac{4 \pi^2}{GP_{3}^2} \times (a'_{12} \sin i')^3=\frac{(M_3 \sin i')^3}{(M_1 + M_2 + M_3)^2} ,
\end{equation}
the corresponding values are displayed in Table 7. The mass function of the third body can be derived as $f(m)=0.00032(\pm0.00003) M_{\odot}$. The mass of the third body $M_3 \sin{i'}=0.087(\pm 0.002) M_{\odot}$ and the orbital radius $a_3$ = 14.33 au. If the minimum mass of the third body is 0.087 $M_{\odot}$, the third body will be a very low-mass star, red dwarf or M-type star that is extremely small luminosity and difficult to detect. Since no third light was reported in the photometric studies (Marino et al. 2007 and Deb \& Singh 2011) and no spectral lines of a third body were found (Rucinski et al. 2007). To check the third body, we also searched for the third light during the photometric solution, but the results showed negative values that means the contribution of the third light is very small comparing to the total light of the system. {\bf Since its minimum mass (i' $\sim$ 90 deg) is small} and locates very far ($\sim$ 14.33 au) from the central binary system, thus it may not play an important role for the origin and evolution of the central binary by removing angular momentum from the inner system via Kozai cycle (Kozai 1962). In this way, RW Dor will normally evolve into contact phase without acceleration.

Based on section 3, there is very weak evidence for cyclic change in $O-C$ analysis of Fig. 3 because of a few eclipse timings before 1948 with large scatter and a big gap of time interval between 1943 and 1980. However, almost contact binaries were found to be triple or multiple systems (Pribulla \& Rucinski 2006; D'Angelo et al 2006; Rucinski 2007) and one can see at Table 6, most of them have no third light in their light curves but show periodic variations in long-term period changes. Recently, a possible substellar object orbiting around the solar-like contact binary V2284 Cyg was firstly reported by Wang et al. (2017). Also in K-type shallow contact binary CC Com with very low mass $M_3 = 0.066M_{\odot}$ by Yang et al. (2009) and KIC 9532219 with $M_3 = 0.089M_{\odot}$ by Lee at el. (2016) or TX Cnc with $M_3 = 0.097M_{\odot}$ by Liu et al. 2007. More recently, the eclipsing binary Kepler-503 has been found that there is a brown dwarf or low-mass star with $M_3 = 0.075M_{\odot}$ orbiting around a subgiant star (Canas et al. 2018). Therefore, the existence of third body in the system cannot be ruled out completely. To prove that the invisible additional companion exists or not, long-term monitoring in photometric with new eclipse timings and spectroscopic observations are required.

\bigskip
This work is supported by the National Natural Science Foundation of China (No. 11503077). We would like to thank Dr. Wiphu Rujopakarn and NARIT for time allocation to use PROMPT-8 for our observations. More CCD observations were obtained with the 2.15-m "Jorge Sahade" (JS) telescope at Complejo Astronomico El Leoncito (CASLEO), San Juan, Argentina. This research has made use of the SIMBAD online database, operated at CDS, Strasbourg, France, NASA's Astrophysics Data System (ADS).

\end{document}